\documentclass[preprint,aps,prb]{revtex4-1} 
\usepackage{graphicx}
\usepackage{natbib}

\begin{document}



\title{Radiation reaction from electromagnetic fields in the neighborhood of a point charge} 
\author{Ashok K. Singal}
\email{ashokkumar.singal@gmail.com}
\affiliation{56, Jugal Apartment, Vastrapur, Ahmedabad - 380 015, India}
\date{\today}
\begin{abstract}
From the expression for the electromagnetic field in the neighborhood of a point charge we determine the rate of electromagnetic momentum 
flow, calculated using the Maxwell stress tensor, across a surface surrounding the charge. From that 
we derive for a `point charge' the radiation reaction formula, which turns out to be 
proportional to the first time-derivative of the acceleration of the charge, identical to the expression for the self-force,   
hitherto obtained in the literature from the detailed mutual interaction between constituents of a 
small charged sphere. We then use relativistic transformations to arrive at a generalized formula for 
radiation reaction for a point charge undergoing a relativistic motion.

\end{abstract}
\maketitle
\section{Introduction}
The radiation reaction, first proposed by Lorentz,~\cite{lor92, lor04} was later derived in detail by Abraham \cite{abr05} and Lorentz~\cite{16} 
from the self-force of an arbitrarily moving small charged sphere. It is in fact 
remarkable that Abraham's formula, \cite{abr05} derived before the advent of the special theory of relativity, 
includes all the relativistic terms correctly. Von Laue  derived the same from a relativistic generalization of Lorentz's non-relativistic formula 
for radiation reaction by expressing it in a 4-vector form. \cite {lau09} An exhaustive treatment of these formulations was later 
provided by Schott \cite{7} and Page and Adams. \cite{24}  The computations of the self-force of a charged sphere have a long history. \cite{20,51} 
These formulas are now available in various forms in modern text-books. \cite{20,1,2,3,25}
In these derivations one usually considers for the charge particle, a classical uniform sphere model 
(a solid ball or a spherical shell) of radius $r_{\rm o}$, which may be made vanishingly small in the limit. One calculates the force 
on an infinitesimal element of the charged sphere due to the fields from all its remainder parts, with the 
positions, velocities and accelerations of the latter calculated at the retarded times.
Then the net force on the charge is calculated by integration over the whole sphere. 
To make it simple, the force calculations are usually done in the instantaneous rest-frame
of the charge where it is generally assumed that the motion of the charged 
varies slowly so that during the light-travel time across the particle, any changes in its
velocity, acceleration and other higher time derivatives
are relatively small. Then one can keep only 
linear terms ${\bf v}/c$, $\dot{\bf v}r_{\rm o}/c^2$, $\ddot{\bf v}r_{\rm o}^2/c^3$, etc., in the formulation of 
self-force, which turns out to be proportional to the rate of change of acceleration. 
Further the self-force is found to be independent of the radius of the small sphere. 
An alternative derivation of the radiation reaction equation was given by Dirac,~\cite{19}  
that involved both retarded and advanced solutions, which is sometimes open to criticism based on the causality
arguments. The problem of radiative losses and its relation with radiation reaction on the charge has been extensively discussed 
in the literature. \cite{pa18,5,6,10,44,ha06,gr09,41,56,57}

The question that we examine here is: 
could the radiation reaction expression, which is independent of the radius of the small sphere, be 
actually derived in the case of a point charge? Of course one cannot meaningfully talk of a self-force on a point charge due to 
electromagnetic interaction between its constituents. 
But there is a possibility to exploit the law of conservation of mechanical and electromagnetic momentum of a system comprising a charge 
and its electromagnetic fields. We shall endeavor to do so here. We accordingly show that the expression 
for radiation reaction derived this way is exactly the same as has been derived for a sphere of vanishingly small radius. 

We shall first examine electromagnetic field in the neighborhood of an arbitrarily moving point charge. From that we 
shall calculate the radiation reaction, for a non-stationary charge moving with a non-relativistic velocity, 
from the electromechanical momentum conservation, using Maxwell stress tensor. Later we will transform the radiation reaction formula 
to a relativistic case. 
\section{Electromagnetic field in the neighborhood of a point charge}
The electromagnetic field (${\bf E},{\bf B}$)  of an arbitrarily moving point charge $e$ at any time $t_{\rm o}$ is given by \cite{1,28}
\begin{equation}
\label{eq:1a}
{\bf E}=\left[\frac{e({\bf n}-{\bf v}/c)}
{\gamma ^{2}r^{2}(1-{\bf n}\cdot{\bf v}/c)^{3}} + 
\frac{e\:{\bf n}\times\{({\bf n}-{\bf v}/c)\times
\dot{\bf v}\}}{rc^2\:(1-{\bf n}\cdot
{\bf v}/c)^{3}}\right]_{ret},
\end{equation}
\begin{equation}
\label{eq:1b}
{\bf B}={\bf n} \times {\bf E}\:,
\end{equation}
where the square bracket indicates that quantities like velocity (${\bf v}$), acceleration ($\dot{\bf v}$), and the 
Lorentz factor ($\gamma=1/\surd(1-(v/c)^{2})$) of the charge are all in terms of their values at the retarded time $t_{\rm o}-r/c$.
The radial vector  ${\bf r}=r{\bf n}$ is from the retarded 
position of the charge to the field point where electromagnetic fields are being evaluated.

Using the vector identity ${\bf v}={\bf n}({\bf v}.{\bf n}) - {\bf n}\times\{{\bf n}\times{\bf v}\}$, 
we can decompose the electric field in Eq.~(\ref{eq:1a}), 
in terms of the radial (along {\bf n}) and transverse components as \cite{17,18}
\begin{eqnarray}
\label{eq:11a}
{\bf E}=\left[\frac{e\:{\bf n}}{\gamma ^2 r^2(1-{\bf n}\cdot{\bf v}/c)^2}
+\frac{e\:{\bf n}\times\{({\bf n}-{\bf v}/c)\times(\gamma{\bf v}+\gamma^3 \dot{\bf v}r/c)\}}
{\gamma^3r^2c\:(1-{\bf n}\cdot{\bf v}/c)^3}\right]_{ret},
\end{eqnarray}
the quantities on the right hand side evaluated 
at the retarded time $t_{\rm o}-r/c$. This is a general expression for the electric field of a charge, 
with the radial (along ${\bf n}$) and transverse (perpendicular to ${\bf n}$) terms fully separated. 

Initially we assume that motion of the charge is non-relativistic and that it 
varies only slowly so that any change in the velocity, acceleration etc. during the light-travel time
across the region of interest are relatively small. This means 
that $|{\bf v}|\,\ll\,c, \;|\dot{{\bf v}}|r/c \,\ll\,c, \;|\ddot{{\bf v}}|r/c \,\ll 
\,|\dot{{\bf v}}|,$ etc. Therefore we drop non-linear terms in ${{\bf v}}$, $\dot{{\bf v}}$ etc., in our formulation.

Thus keeping only linear terms in ${\bf v}$ and $\dot{\bf v}$, from Eq.~(\ref{eq:11a}) we can write 
\begin{equation}
\label{eq:12h}
{\bf E}=e\:{\bf n}\bigg[\frac{1}{r^{2}}+\frac{2{\bf n} \cdot {\bf v}}{r^{2}c}\bigg]_{ret}
+e\:{\bf n}\times \bigg[\frac{{\bf n}\times ({\bf v}+\dot{\bf v}r/c)}{r^{2}c}\bigg]_{ret} \;.
\end{equation}

Now, under the above conditions, one can make a Taylor series expansion
of the retarded velocity and acceleration vectors around the present time $t_{\rm o}$ 
\begin{equation}
\label{eq:1d}
{{\bf v}}={{\bf v}}_{\rm o}-\frac{\dot{{\bf v}}_{\rm o}r}{c}+\frac{\ddot{{\bf v}}_{\rm o}r^{2}}{2c^{2}}+\cdots,
\end{equation}
\begin{equation}
\label{eq:1f}
\dot{{\bf v}}=\dot{{\bf v}}_{\rm o}-\frac{\ddot{{\bf v}}_{\rm o}r}{c}+\cdots,
\end{equation}
all vector quantities on the right hand side evaluated at the present time $t_{\rm o}$. 

Substituting for ${\bf v}$ and $\dot{\bf v}$ from Eqs.~(\ref{eq:1d}) and (\ref{eq:1f}) into 
Eq.~(\ref{eq:12h}), and dropping terms of order $r$ or its higher powers, that will become zero as $r\rightarrow 0$, 
we get for the electric field in the neighborhood of an accelerated point charge as
\begin{eqnarray}
\label{eq:12i}{\bf E}=e\:{\bf n}\bigg[\frac{1}{r^{2}}
+\frac{2{\bf n} \cdot {\bf v}_{\rm o}}{r^{2}c}-\frac{2{\bf n} \cdot \dot{\bf v}_{\rm o}}{r c^2}
+\frac{{\bf n} \cdot \ddot{\bf v}_{\rm o}}{c^{3}}\bigg] 
+\: e\:{\bf n}\times \left[\frac{{\bf n}\times {\bf v}_{\rm o}}{r^{2}c}-
\frac{{\bf n}\times \ddot{\bf v}_{\rm o}}{2c^{3}}\right],
\end{eqnarray}
and the magnetic field as
\begin{equation}
\label{eq:12k}
{\bf B}= e\left[-\frac{{\bf n}\times {\bf v}_{\rm o}}{r^{2}c}+\frac{{\bf n}\times \ddot{\bf v}_{\rm o}}{2c^{3}}\right].
\end{equation}
It should be noted that in Eqs.~(\ref{eq:12i}) and (\ref{eq:12k}), while the radial distance $r$ and the unit vector ${\bf n}$ 
are specified with respect to the retarded position of the charge, the velocity and acceleration at the retarded time 
have been expressed in terms of the charge motion at the present time $t_{\rm o}$ through Eqs.~(\ref{eq:1d}) and (\ref{eq:1f}). 
\section{The computation of radiation reaction from Maxwell stress tensor}
Law of conservation of momentum in an electromagnetic system, containing both charges and fields, states that 
at any given instant, say $t_{\rm o}$, the rate of change of the combined momentum in a volume $V$, comprising the mechanical momentum of the charges 
$({\bf P}_{\rm mech})$ and the electromagnetic momentum  of the fields (${\bf P_{\rm field}}$), is equal to the rate of momentum flow into 
the volume through a surface $\Sigma$ surrounding that volume. \cite{1,25}
\begin{equation}
\label{eq:1f1}
\frac{{\rm d}({\bf P}_{\rm mech}+{\bf P_{\rm field}})}{{\rm d}t}=\oint_{\Sigma}{{\rm d}\Sigma}\;{\bf n}\; \cdot \stackrel{\leftrightarrow}{\bf T}\:,
\end{equation}
where all quantities are calculated for the same instant of time $t_{\rm o}$ . 
The vector component ${\bf n} \;\cdot \stackrel{\leftrightarrow}{\bf T}$ of the Maxwell stress tensor $\stackrel{\leftrightarrow}{\bf T}$, represents the radial momentum flow into 
the volume per unit area of the surface, and can be calculated from \cite{1,25}
\begin{equation}
\label{eq:1f1a}
{\bf n} \;\cdot \stackrel{\leftrightarrow}{\bf T}\;=\frac{1}{4\pi}\bigg[({\bf n} \cdot {\bf E}){\bf E}+({\bf n} \cdot {\bf B}){\bf B}
- \frac{{\bf n}(E^2+B^2)}{2}\bigg]. 
\end{equation}
The volume integral of the electromagnetic field momentum enclosed within the spherical surface $\Sigma$ is given by
\begin{equation}
\label{eq:1f1b}
{\bf P_{\rm field}}=\frac{1}{4\pi c}\int{{\rm d}V}\:({\bf E}\times{\bf B}).
\end{equation}
\begin{figure}[t]
\includegraphics[width=\columnwidth]{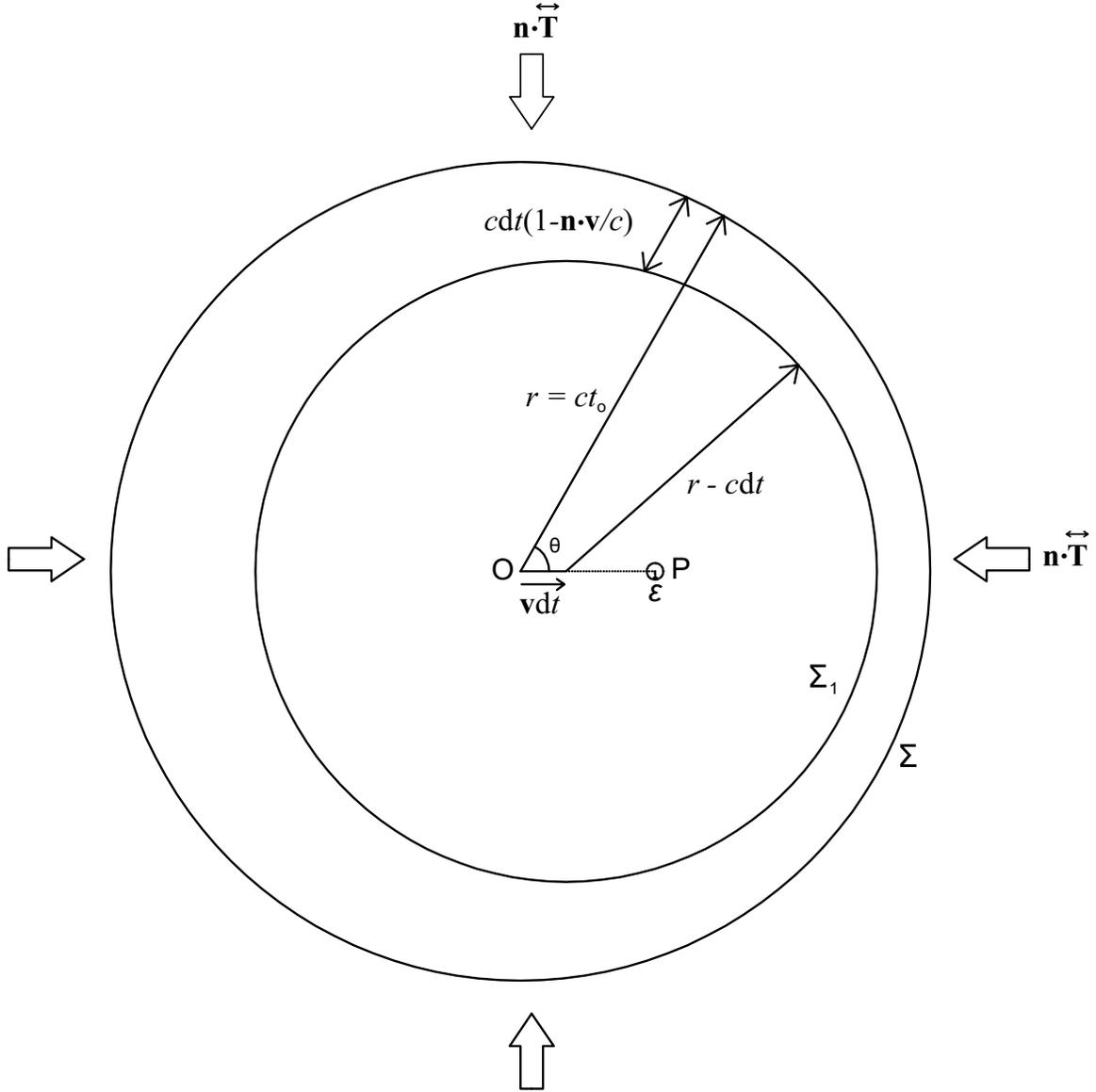}
\caption{Geometry for calculating the volume element in the case of a moving charge. The spacing between the two spheres in direction $\bf n$ is 
$dr(1-{\bf n}\cdot {\bf v} /c)$. The vector component of the Maxwell stress tensor 
${\bf n} \; \cdot \stackrel{\leftrightarrow}{\bf T}$
represents the radial electromagnetic momentum flow into the volume through a unit area of the surface $\Sigma$.}
\end{figure}

Let us consider a charge moving with an arbitrary but non-relativistic speed ($v \ll c$). 
We assume that the moving charge was at time $t=0$ at a point O, and now at time $t_{\rm o}$ the charge is at its `present' position P.
Let $\Sigma$ be a spherical surface of radius $r=ct_{\rm o}$, centered on O and let $V$ be the volume enclosed within (Fig. 1). 
The electromagnetic field at any point on surface $\Sigma$ is 
determined from the position and motion of charge at O, which is the corresponding time-retarded position of the charge. 
We want to use the law of momentum conservation to compute the mechanical force on the charge at P.
It should be point out here that although the present position of the charge is not at the center of the sphere $\Sigma$, 
yet the charge and its neighborhood is still enclosed well within the surface $\Sigma$ (as $v \ll c$) 
and the electromagnetic momentum conservation is as much applicable to the electromagnetic momentum flux through 
this surface as to a surface centered on the charge. As the charge moves, different spherical surfaces within the volume are causally 
related to different retarded positions of the charge, and the corresponding  
radial vector $\bf r$ and the unit vector $\bf n$ are constantly changing. In Fig. 1, we have shown another spherical surface $\Sigma_1$ of radius 
$r-{\rm d}r=c(t_{\rm o}-{\rm d}t)$, centered at the corresponding time-retarded position of the charge at a distance ${\bf v} {\rm d}t$ from O.
In order to calculate the volume integral of the field  momentum, we first evaluate the volume element between the two 
spherical surfaces $\Sigma$ and $\Sigma_{1}$, centered
on two different retarded positions of the charge, separated by $\bf v{\rm d}t$  (Fig.~1). The radial 
distance between the two spherical surfaces is not a constant $dr$ but is instead $dr(1-{\bf n}\cdot {\bf v} /c)$, with the corresponding 
volume element given as $dV=2\pi r^{2} (1-(v/c) \cos \theta)\sin \theta \,dr \,d\theta$.
Using Eqs.~(\ref{eq:12i}) and (\ref{eq:12k}), and dropping all terms which are non-linear in $\bf v_{\rm o}$ or its derivatives, or those 
which upon integration over the solid angle of the whole sphere yield a zero net value, or those proportional to $r$ or its higher powers 
which would vanish when $r\rightarrow 0$, we get 
\begin{equation}
\label{eq:1f2}
{\bf P_{\rm field}}=\frac{-e^2}{4\pi c}\int{{\rm d}V}\;\frac{{\bf n}\times({\bf n}\times {\bf v}_{\rm o})}{r^4c}
=\frac{2e^2{\bf v}_{\rm o}}{3\: c^{2}}\int_{\epsilon}^{r}\frac {{\rm d}r}{r^2}
=\frac{2e^2{\bf v}_{\rm o}}{3\: c^{2}}\left[\frac {1}{\epsilon}-\frac {1}{r}\right].
\end{equation}
This is the electromagnetic field momentum in the volume $V$, enclosed within the spherical surface $\Sigma$, at time $t_{\rm o}$. 
The lower limit of the integral is restricted to $\epsilon$, as done even in static Coulomb fields to avoid divergence of the self-field energy 
at low $r$. Then we get
\begin{equation}
\label{eq:1f3}
\frac{{\rm d}{\bf P_{\rm field}}}{{\rm d}t}=\frac{2e^2\dot{\bf v}_{\rm o}}{3\: c^{2}}\left[\frac {1}{\epsilon}-\frac {1}{r}\right].
\end{equation}
The first term on the right hand side in Eq.~(\ref{eq:1f3}) represents the change in momentum of the self-fields of the charge 
as its velocity changes. Its divergence, as $\epsilon\rightarrow 0$, is unavoidable due to the `point' nature of the considered charge distribution 
where the Coulomb field energy of even a stationary charge diverges. The formulation for a point charge remains arbitrary 
to that extent. \cite{p32.1} 

We substitute for electric and magnetic fields from Eqs.~(\ref{eq:12i}) and (\ref{eq:12k}) into Eq.~(\ref{eq:1f1a}), and integrate over the 
surface $\Sigma$ to obtain the electromagnetic momentum flow into enclosed volume $V$ at $t_{\rm o}$ as
\begin{eqnarray}
\nonumber
\oint_{\Sigma}{{\rm d}\Sigma}\;{\bf n} \;\cdot \stackrel{\leftrightarrow}{\bf T}\;&= &\frac{e^2}{4\pi}\oint_{\Omega} r^{2} {\rm d}\Omega
\bigg[\frac{\bf n}{r^{2}}\bigg\{\frac{2{\bf n} \cdot {\bf v}_{\rm o}}{r^{2}c}-\frac{2{\bf n} \cdot \dot{\bf v}_{\rm o}}{r c^2}
+\frac{{\bf n} \cdot \ddot{\bf v}_{\rm o}}{c^{3}}\bigg\}\\
\label{eq:1d2}
&&\;\;\;\;\;\;\;\;\;\;\;\;\;\;\;\;\;\;\;\;+\frac{\bf n}{r^{2}}\times \left\{\frac{{\bf n}\times {\bf v}_{\rm o}}{r^{2}c}-
\frac{{\bf n}\times \ddot{\bf v}_{\rm o}}{2c^{3}}\right\}\bigg],
\end{eqnarray}
where $\Omega$ is the solid angle. Upon integration, terms containing ${\bf v}_{\rm o}$ get cancelled, leaving us with
\begin{eqnarray}
\nonumber
\oint_{\Sigma}{{\rm d}\Sigma}\;{\bf n} \;\cdot \stackrel{\leftrightarrow}{\bf T}\;&= &\frac{e^2}{4\pi}\oint_{\Omega} {\rm d}\Omega
\bigg[-\frac{2{\bf n}({\bf n} \cdot \dot{\bf v}_{\rm o})}{r c^2}+\frac{{\bf n}({\bf n} \cdot \ddot{\bf v}_{\rm o})+\ddot{\bf v}_{\rm o}}{2c^3}\bigg]\\
\nonumber
&=& \frac{e^2}{2}\int_{\rm o}^{\pi} {\rm d}\theta\:\sin\theta\bigg[-\frac{2\dot{\bf v}_{\rm o}\cos^2\theta}{r c^2}
+ \frac{\:\ddot{\bf v}_{\rm o} \cos^2\theta+\ddot{\bf v}_{\rm o} }{2 c^3}\bigg]\\
\label{eq:1d4}
&=&-\frac{2e^{2}}{3r c^{2}}\dot{\bf v}_{\rm o}+ \frac{2e^{2}}{3 c^{3}}\ddot{\bf v}_{\rm o}\:.
\end{eqnarray}
Now from Eq.~(\ref{eq:1f1}), the mechanical force ${\bf F}={{\rm d}{\bf P}_{\rm mech}}/{{\rm d}t}$, on the charge enclosed 
within $V$ at $t_{\rm o}$ is
\begin{equation}
\label{eq:1f1c}
{\bf F}=\int_{\Sigma}{{\rm d}\Sigma}\;{\bf n}\; \cdot \stackrel{\leftrightarrow}{\bf T}-\frac{{\rm d}{\bf P_{\rm field}}}{{\rm d}t}\:.
\end{equation}
Substituting Eqs. (\ref{eq:1f3}) and (\ref{eq:1d4}) into Eq.~(\ref{eq:1f1c}), we get the electromagnetic force on 
the charge at time $t_{\rm o}$ as
\begin{equation}
\label{eq:1d5}
{\bf F}=-\frac{2e^{2}}{3\:\epsilon \: c^{2}}\dot{\bf v}_{\rm o}+ \frac{2e^{2}}{3 c^{3}}\ddot{\bf v}_{\rm o}\:,
\end{equation}
where $\dot{\bf v}_{\rm o}$ and $\ddot{\bf v}_{\rm o}$ respectively are the acceleration and the rate of change of acceleration of the 
charge at $t_{\rm o}$.
An interesting thing to note here is that one could have arrived at Eq.~(\ref{eq:1d5}), without evaluating the field momentum  
(Eqs.~(\ref{eq:1f2}) and (\ref{eq:1f3})), by using $\epsilon$ as the radius of the spherical surface $\Sigma$ directly in Eq.~(\ref{eq:1d4}). 
However, from Eq.~(\ref{eq:1f2}) we learn that $2e^2{\bf v}_{\rm o}/(3\: c^2 \epsilon)$ is the field momentum in the volume exterior to 
a radial distance $\epsilon$ from 
the point charge (we let $r\rightarrow \infty$ in the last term in Eq.~(\ref{eq:1f2})), due to its present velocity ${\bf v}_{\rm o}$. This is true for 
any value of $\epsilon$. This of course is a well-known result for a uniformly moving charge, \cite{29,15} but now we know that 
this formula is applicable even if the charge were moving arbitrarily but with a non-relativistic motion. As the charge velocity changes due to its 
acceleration, the field momentum, $\propto {\bf v}_{\rm o}$, also must accordingly change. Therefore, from the functional form of the first term on 
the right hand side of  Eq.~(\ref{eq:1d5}), it is evident that this term is due to nothing else but the changing electromagnetic momentum in the fields 
because of the changing velocity of the charge and has nothing to do directly with the radiative losses by the charge. It is only the second term in the 
force formula Eq.~(\ref{eq:1d5}), independent of $r$, which represents radiation reaction and is exactly the same as obtained from a derivation 
of the self-force of an accelerated charged sphere of small dimensions. \cite{16,7,24,1,2,3,25,20}  

Thus the expression for radiation reaction
\begin{equation}
\label{eq:1d6}
{\bf F}=\frac{2e^{2}}{3 c^{3}}\ddot{\bf v}_{\rm o}\:,
\end{equation}
being proportional to rate of change of acceleration, has been derived here from the Maxwell stress tensor of a {\em point charge}. 

From the Poynting flux in the neighborhood of a point charge moving with a non-relativistic but otherwise arbitrary velocity ${\bf v}_{\rm o}$, 
it has recently been shown \cite{68a} that the radiative power loss is given by 
\begin{equation}
\label{eq:1s}
{\cal P}= \frac{-2e^2\ddot{\bf v}_{\rm o}\cdot{\bf v}_{\rm o}}{3c^{3}}\;,
\end{equation}
which, using Eq.~(\ref{eq:1d6}), can be written as
\begin{equation}
\label{eq:1t}
{\cal P}=  -{\bf F}\cdot{\bf v}_{\rm o}\;.
\end{equation}
This is consistent with ${\bf F}$ being a radiative drag force on the moving charge.
\section{Radiation Reaction for a Relativistic Motion of the Charge}
We calculated the radiation reaction (Eq.~(\ref{eq:1d6})) for a point charge having a non-relativistic 
but otherwise arbitrary motion. We shall now use relativistic transformations to obtain expression for radiation reaction for a 
charge moving with relativistic speeds. 

In an inertial frame ${\cal K}$ (lab-frame!) let the charge at some instant $t$ be moving with velocity ${\bf v}$ 
and acceleration $\dot{\bf v}$ and let at that event ${\cal K}'$ be the instantaneously rest frame of the charge, thus moving 
with a velocity ${\bf v}_\parallel={\bf v}$ with respect to ${\cal K}$. The proper time 
interval ${\rm d}t'$ in rest frame ${\cal K}'$ is related to the corresponding time interval in the lab-frame ${\cal K}$ 
as ${\rm d}t=\gamma{\rm d}t'$. 

In the instantaneous rest frame ${\cal K}'$ we rewrite radiation reaction (Eq.~(\ref{eq:1d6})) as
\begin{equation}
\label{eq:1d6a}
{\bf F}'=\frac{2e^{2}}{3 c^{3}}\ddot{\bf v}'\:,
\end{equation}
where we have dropped the subscript from ${\bf v}_{\rm o}$ and its temporal derivatives. 
We now want to transform this relation to the frame ${\cal K}$.

The force components transform from the rest frame ${\cal K}'$ to the lab-frame ${\cal K}$ as \cite{71} 
\begin{equation}
\label{eq:3f2.3}
{\bf F}'_\parallel={\bf F}_\parallel\:,
\end{equation}
\begin{equation}
\label{eq:3f2.4}
{\bf F}'_\perp = \gamma{\bf F}_\perp \:.
\end{equation}

The parallel component of acceleration transforms as
\begin{equation}
\label{eq:3f2.1}
\dot{\bf v}'_\parallel = \gamma^3 \dot{\bf v}_\parallel \:,
\end{equation}
from which we get 
\begin{equation}
\label{eq:3f2.1a}
\ddot{\bf v}'_\parallel = \gamma(3 \gamma^5 (\dot{\bf v}\cdot{\bf v}/c^2)\dot{\bf v}_\parallel 
+\gamma^3 \ddot{\bf v}_\parallel)\:,
\end{equation}
 
On the other hand the perpendicular component of acceleration transforms as \cite{71} 
\begin{equation}
\label{eq:3f2.2}
\dot{\bf v}'_\perp = \gamma^2 \dot{\bf v}_\perp+\gamma^4 (\dot{\bf v}\cdot{\bf v}/c^2){\bf v}_\perp\:.
\end{equation}
From that we get
\begin{equation}
\label{eq:3f2.2a}
\ddot{\bf v}'_\perp = 2 \gamma^5 (\dot{\bf v}\cdot{\bf v}/c^2) \dot{\bf v}_\perp+\gamma^3 \ddot{\bf v}_\perp
+\gamma^5 (\dot{\bf v}\cdot{\bf v}/c^2)\dot{\bf v}_\perp\:.
\end{equation}
It should be noted that though the instantaneous value ${\bf v}_\perp=0$ (by the definitions of frames ${\cal K}'$ and ${\cal K}$),
we retained it in Eq.~(\ref{eq:3f2.2}) as it is needed in the derivation of Eqs.~(\ref{eq:3f2.2a}), where  $\dot{\bf v}_\perp$ 
need not be zero.

Thus for the parallel component of force from Eqs.~(\ref{eq:1d6a}), (\ref{eq:3f2.3}) and (\ref{eq:3f2.1a}) we get  
\begin{equation}
\label{eq:3f1}
{\bf F}_\parallel= \frac{2e^2}{3c^3}\left[\frac{3\gamma ^{6}(\dot{\bf v}\cdot{\bf v})\dot{\bf v}_\parallel}{c^{2}}
+\gamma ^{4}\ddot{\bf v}_\parallel\right] \:,
\end{equation}
while for the perpendicular component of force Eqs.~(\ref{eq:1d6a}), (\ref{eq:3f2.4}) and (\ref{eq:3f2.2a}) lead us to 
\begin{eqnarray}
\label{eq:3f5}
{\bf F}_\perp = \frac{2e^{2}}{3 c^{3}} \left[\frac{3\gamma^{4}(\dot{\bf v}\cdot{\bf v})\dot{\bf v}_\perp}{c^{2}}
+\gamma^2 \ddot{\bf v}_\perp\right] \:.
\end{eqnarray}
Combining the two force components and after a rearrangement of various terms we get the required relativistic expression 
for the radiation reaction as
\begin{eqnarray}
\label{eq:3i}
{\bf F}= \frac{2e^{2}\gamma^2}{3 c^{3}}\left[ \ddot{\bf v}+\frac{\gamma ^{2}(\ddot{\bf v}\cdot{\bf v}){\bf v}}{c^{2}}
+\frac{3\gamma ^{2}(\dot{\bf v}\cdot{\bf v})\dot{\bf v}}{c^{2}}+\frac{3\gamma ^{4}(\dot{\bf v}\cdot{\bf v})^2{\bf v}}{c^{4}}\right].\:
\end{eqnarray}
which is the relativistic form of the radiation reaction formula, earlier derived in literature by various different methods. \cite{abr05,lau09,7,24,20} 

From the relativistic expression (Eq.~(\ref{eq:3i})) for the radiative drag force, the radiative power loss in a relativistic case is
\begin{equation}
\label{eq:10}
{\cal P}= -{\bf F}\cdot{\bf v}=-\frac{2e^{2}\gamma ^{4}}{3c^{3}}\left[\ddot{\bf v}\cdot{\bf v}+
3\gamma ^{2}\frac{(\dot{\bf v}\cdot{\bf v})^{2}}{c^{2}}\right]\:.
\end{equation}
\section{Conclusions}
From the electromagnetic field in the neighborhood of a `point charge' in arbitrary motion, we derived here the 
radiation reaction from the rate of electromagnetic momentum flow, calculated using the Maxwell stress tensor, across a surface surrounding 
the point charge. Our result that the radiation reaction 
is proportional to the first time-derivative of the acceleration of the charge, is in complete 
agreement with the self-force of a small charged sphere, determined in literature from a  
detailed mutual interaction between its constituents. We also derived a generalized formula for 
radiation reaction for a relativistic motion of the point charge, using relativistic transformations.

{}
\end{document}